\documentclass[a4paper,preprintnumbers,floatfix,superscriptaddress,aps,pra,twocolumn,showpacs,notitlepage,longbibliography]{revtex4-1}
\usepackage[utf8]{inputenc}
\usepackage[T1]{fontenc}
\usepackage[sc,osf]{mathpazo}
\usepackage{amsmath, amsthm, amssymb,amsfonts,mathbbol,amstext}
\usepackage{graphicx}
\usepackage{dcolumn}
\usepackage{bm}
\usepackage{bbm}
\usepackage[colorlinks=true,linkcolor=blue,urlcolor=blue,citecolor=blue,pdfusetitle]{hyperref}
\usepackage{mathtools}
\usepackage{comment}
\usepackage{color}
\usepackage{dsfont}
\usepackage[normalem]{ulem}
\usepackage{multirow}
\usepackage{diagbox}
\usepackage{pdfcomment}

\newcommand\norm[1]{\left\lVert#1\right\rVert}

\newcommand{\bra}[1]{\langle #1|}
\newcommand{\ket}[1]{|#1\rangle}

\newcommand{\ba}{\begin{eqnarray}}
\newcommand{\be}{\begin{equation}}
\newcommand{\ee}{\end{equation}}
\newcommand{\ea}{\end{eqnarray}}
\newcommand{\ban}{\begin{eqnarray*}}
\newcommand{\ean}{\end{eqnarray*}}

\definecolor{forestgreen}{rgb}{0.13, 0.55, 0.13}

{\begin{framed}\begin{small}}
{\end{small}\end{framed}}

\DeclareGraphicsExtensions{.pdf,.png,.jpg}

\makeindex

\begin{document}

\title{Enhancing entanglement and total correlations dynamics via local unitaries}

\author{Joab Morais Varela}
\affiliation{Departamento de Física Teórica e Experimental, Universidade Federal do Rio Grande do Norte, Natal, RN, 59078-970, Brazil}
\author{Ranieri Nery}
\affiliation{International Institute of Physics, Federal University of Rio Grande do Norte, 59070-405 Natal, Brazil}
\author{George Moreno}
\affiliation{International Institute of Physics, Federal University of Rio Grande do Norte, 59070-405 Natal, Brazil}
\author{Alice Caroline de Oliveira Viana}
\affiliation{Departamento de Física Teórica e Experimental, Universidade Federal do Rio Grande do Norte, Natal, RN, 59078-970, Brazil}
\author{Gabriel Landi}
\affiliation{Instituto de F\'isica da Universidade de S\~ao Paulo,  05314-970 S\~ao Paulo, Brazil.}
\author{Rafael Chaves}
\affiliation{International Institute of Physics, Federal University of Rio Grande do Norte, 59070-405 Natal, Brazil}
\affiliation{School of Science and Technology, Federal University of Rio Grande do Norte, 59078-970 Natal, Brazil}

\begin{abstract}
The interaction with the environment is one of the main obstacles to be circumvented in practical implementations of quantum information tasks. The use of local unitaries, while not changing the initial entanglement present in a given state, can enormously change its dynamics through a noisy channel, and consequently its ability to be used as a resource. This way, local unitaries provide an easy and accessible way to enhance quantum correlations in a variety of different experimental platforms. Given an initial entangled state and a certain noisy channel, what are the local unitaries providing the most robust dynamics?  In this paper we solve this question considering two qubits states, together with paradigmatic and relevant noisy channels, showing its consequences for  teleportation protocols and identifying cases where the most robust states are not necessarily the ones imprinting the least information about themselves into the environment. We also derive a general law relating the interplay between the total correlations in the system and environment with their mutual information built up over the noisy dynamics. Finally, we employ the IBM Quantum Experience to provide a proof-of-principle experimental implementation of our results.
\end{abstract}

\maketitle

\section{Introduction}

Considerable theoretical and experimental advances have been achieved in the quantum control of a variety of systems, leading recently to the first examples of quantum supremacy \cite{Arute2019, Zhong1460,wu2021strong}. That is, a quantum based protocol outperforming,  by of orders of magnitude, the efficiency of existing classical platforms. These examples, however, solve problems of limited applicability and, more importantly, are not fault-tolerant \cite{Knill1998} or error-corrected \cite{lidar2013quantum}. Indeed, any operational universal quantum computer will unavoidably need to counteract the detrimental effects of decoherence and faulty logical operations, which would otherwise accumulate exponentially fast, and washout any quantum advantage \cite{lidar2013quantum,Knill1998, Aharonov2008}. The use of fault-tolerant quantum error-correction mechanisms provide a reliable implementation of quantum devices \cite{Preskill1998}. However, such mechanisms introduce enormous overhead in the components that, moreover, require a precise control over each of the constituents and are still in its infancy \cite{cory1998experimental,yao2012experimental,andersen2020repeated}.

Fortunately, as illustrated by the recent quantum supremacy experiments, a wide variety of particular computational and communication tasks with quantum advantages and, more importantly, without the need of error-correction codes, have been identified. Quantum correlations and, in particular, entanglement have been recognized over the last two
decades as the key resource in a variety of physical tasks \cite{Horodecki2009}. These range from
teleportation \cite{Bennet1993}, dense-coding and communication-complexity \cite{Buhrman2010}, to metrology protocols \cite{Giovannetti2006} and randomness certification \cite{acin2016certified}. For some tasks the detrimental
effects of decoherence can be minimized, and sometimes even ignored, without resorting to any complex fault tolerant
error correction scheme. For example, in a teleportation protocol, bipartite quantum systems provide gain over their
classical counterparts whenever the quantum state is entangled \cite{Horodecki1999}. Using this, faithful quantum teleportation with
$80\%$ accuracy, far beyond the classical limit of $67\%$, has been experimentally realized over distances of up to 1400 Km \cite{ren2017ground}.

For fragile quantum states, a possible way to enhance the robustness of their entanglement and other quantum properties is to properly choose their local encoding prior to the action of noise. For instance, while local unitaries do not change the entanglement of the initial state, the situation is dramatically changed for its dynamics through a noisy channel \cite{Hein2005,Ziman2006,aolita2008scaling,cavalcanti2009open,Chaves2012,chaves2012multipartite,chaves2013noisy}. Interestingly, the enhancement is achieved passively, without having to perform any error-syndrome. And, more importantly, with no cost at all in terms of extra qubits. The local encoding alone guarantees a high robustness of the entanglement. In spite of its simplicity and applicability to many current experimental platforms \cite{proietti2019enhanced}, the quantum enhancement -- be it on entanglement or other forms of correlations \cite{Modi2012,Goold_2015,Landi2020a}-- provided by local unitaries is still fairly unexplored even in the two-qubit case. For a certain initial state and a given noisy channel, what is the local unitarily-equivalent state preserving most of its correlations through the decoherent dynamics? 

This is the question we address in this paper. Considering generic initial states and noisy dynamics, we first show a general law relating the mutual information between the system and environment with the changes in their total correlations \cite{Modi2012,Goold_2015}. Following that we consider what are the optimal unitaries that can be applied to a given two-qubit entangled state in order to preserve, as best as possible, its entanglement through the dynamics. We obtain analytical answers to specific classes of states and channels, also showing its consequences in the use of these noisy states as a resources in a quantum teleportation protocol. Surprisingly, for some choices of initial entangled states and channels, we observed a counter-intuitive phenomenon: the most robust state against noise is exactly that which is most entangled with the environment. Finally, by employing the IBM Quantum Experience, we also test experimentally the enhancement provide by local unitaries. 

The paper is organized as follows. In Sec. \ref{sec:sec2} we introduce the quantum channels considered in this work.  In Sec. \ref{sec:IBMQ} we describe our implementation in the IBM Q-Experience. In Sec. \ref{sec:sec3} we discuss the measure of correlations introduced in~\cite{Modi2012,Goold_2015}, derive a general law connecting such measure of correlation with the system-environment mutual information generated through the noisy dynamics. In Sec. \ref{sec:sec4} we consider different entanglement measures and analytically show what are the most and least robust two-qubit entangled states, and a number of its consequences. In Sec. \ref{sec:sec5} we discuss our findings and point out interesting directions for future research.

\section{Open quantum systems dynamics}
\label{sec:sec2}
In what follows, we consider a system of $N$ qubits, prepared in a generic, globally correlated state $\rho_S$.
Each qubit is then individually coupled to its own local environment. This represents a typical situation, where a source distributes subsystems to distant parties. 
The dynamics of the $i$-th qubit is governed by 
a completely positive trace-preserving map
$\Lambda_i$ that can be described in terms of Kraus operators $K_{ij}$, such that the evolution of the i-th subsystem $\rho_i$ is given by $\Lambda_i(\rho_i)=\sum_{j} K_{ij} \rho_i K^{\dagger}_{ij}$. The strength of the noise and the time it acts are conveniently parametrized by a variable $p \in [0,1]$, such that $p=0$ for $t=0$ and $p \rightarrow 1$ for $t \rightarrow \infty$ . Thus, the joint noisy state after time $p$ is obtained by the composition of the individual evolutions $\Lambda_i$, i.e., $\rho_S(p) = \Lambda(\rho_S):= \Lambda_{1}\otimes ...\otimes
\Lambda_{N}(\rho_S)$.

Here we will be mainly interested in two paradigmatic noisy channels: dephasing and  amplitude damping. The dephasing channel (D) has Kraus operators given by
\begin{equation}
K_0=\sqrt{1 - \frac{p}{2}}\begin{pmatrix}
1 & 0 \\
0 & 1
\end{pmatrix},
\quad K_1=\sqrt{\frac{p}{2}}\begin{pmatrix}
1 & 0 \\
0 & -1
\end{pmatrix},
\end{equation}
and represents the situation in which, without any energy exchange, there is loss of quantum information with probability p. This channel describes many experiments with atomic or ionic qubits, where the dominant source of noise is
dephasing from e.g. magnetic-field or laser-intensity fluctuations \cite{leibfried2003quantum}.

In turn, the amplitude damping (AD) has Kraus operators are given by
\begin{equation}
K_0= \begin{pmatrix}
1 & 0 \\
0 & \sqrt{1-p}
\end{pmatrix},
\quad K_1=\begin{pmatrix}
0 & \sqrt{p} \\
0 & 0
\end{pmatrix},
\end{equation}
and represents a purely dissipative process, describing the relaxation from the excited state to the ground state, as in spontaneous emission.

For our purposes it will also be relevant to analyse the joint system-environment quantum state. 
Without loss of generality, we will assume that each local environment is also composed of a single qubit.
The initial state is taken to be pure, and factorized, as $\rho_{SE}=\rho_S \otimes \rho_E$. Moreover, we take $\rho_E=\left( \ket{0}\bra{0} \right)^{\otimes N}$. Since each subsystem $\rho_i$ is coupled locally to its own environment, they evolve according to a global unitary such that $\Lambda_i(\rho_i)=\mathrm{Tr}_{E_i} \left[ V_i (\rho_i \otimes \ket{0}\bra{0}) V^{\dagger}_i\right]$. In this unitary evolution description, the dephasing channel is described by the map
\begin{eqnarray}
\ket{0}_S\ket{0}_E \rightarrow \sqrt{1-p/2}\ket{0}_S\ket{0}_E+\sqrt{p/2}\ket{0}_S\ket{1}_E, \\ \nonumber
\ket{1}_S\ket{0}_E \rightarrow \sqrt{1-p/2}\ket{1}_S\ket{0}_E-\sqrt{p/2}\ket{1}_S\ket{1}_E.
\end{eqnarray}
In turn, the amplitude damping channel is given by
\begin{eqnarray}
\ket{0}_S\ket{0}_E &\rightarrow& \ket{0}_S\ket{0}_E, \\ \nonumber
\ket{1}_S\ket{0}_E &\rightarrow& \sqrt{1-p}\ket{1}_S\ket{0}_E+\sqrt{p}\ket{0}_S\ket{1}_E.
\end{eqnarray}

\section{Experimental Implementation in the IBM Q-Experience}
\label{sec:IBMQ}
For the experimental implementation we have used the online platform IBM Q-experience \cite{IBMQX, Dwyer2020} that offers access to real quantum computers. In particular we have used a 5-qubits computer with 32 quantum volume \verb|ibmq_manila|. To create the circuits we used the open-source SDK Qiskit \cite{gadi2019}, which allows one to create, compile and run quantum circuits, on the desired platform; it also provides tools to analyze and post-process the experimental data.

\begin{figure} [!t]
\centering
\includegraphics[width=0.7\linewidth]{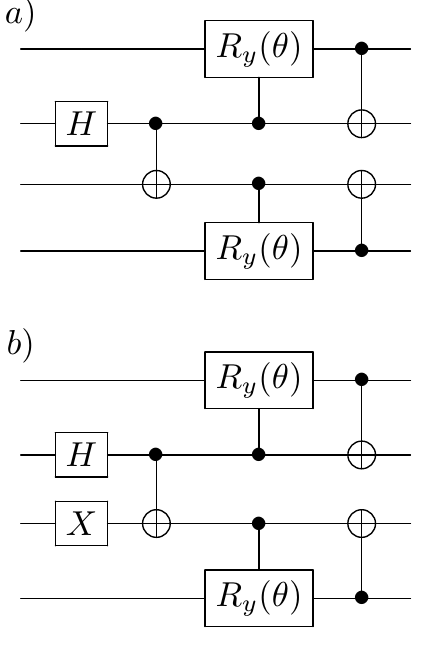}
\caption {\textbf{Circuits implementing the amplitude damping channel.}
\textbf{a)} For the state 
$|\Phi_{0} \rangle =(1/\sqrt{2})\left( \ket{00}+\ket{11} \right)$. 
\textbf{b)} For the state 
$|\Phi_{\pi / 2} \rangle=(1/\sqrt{2})\left( \ket{01}+\ket{10} \right)$. All the qubits in the circuit are initialized as $|0\rangle$. Notice that the single difference between both circuits is an $X$ gate applied in the beginning of the circuit. As we will explore in details, this is already enough to enhance substantially the dynamics of the correlations in the system of interest. } \label{fig:circuits}
\end{figure}

In this paper, we simulate the open system dynamics by implementing the AD channel on the two-qubit entangled states, $|\Phi_{0} \rangle =(1/\sqrt{2})\left( \ket{00}+\ket{11} \right)$ and $|\Phi_{\pi / 2} \rangle=(1/\sqrt{2})\left( \ket{01}+\ket{10} \right)$. 
This is accomplished with the circuits shown in Fig.\ref{fig:circuits}. Note that the decoherence factor is controlled by the rotation gate $R_{y}(\theta)$, with $p = sin^2(\frac{\theta}{2})$.

To obtain the experimental density matrix we use Qiskit's tools for performing quantum state tomography. For that, measurements corresponding to all Pauli-basis combinations are implemented, with a total of 8192 shots for each combination. The final state is obtained by a a maximum-likelihood estimation state tomography fitter \cite{Smolin2012}. Due to external noise effects, that cause a wrong readout output, it is crucial to perform an error mitigation process. Once again, we employ a template tool provided by Qiskit, which consists of performing a measurement calibration given by a list of circuits that prepare each basis state. 

With the density matrix at hand we can evaluate any function of it, in particular compute the entanglement and correlation dynamics of the system-environment. To reduce the unavoidable fluctuations of the IBM quantum computer, we realize each tomography multiple times
and average over these realizations when computing a given entanglement or correlation measure.

\section{Total correlations and system-environment correlations}
\label{sec:sec3}

As the system of interest interacts with the environment, they become entangled with each other. The quantum information and correlations originally contained in the system spread over the joint system-environment quantum state. In this context, our first result will be an equation connecting the flow of correlations from system to environment, as correlations built between them.

As a benchmark to analyze the flow of correlations from system to the environment we will employ the so-called total correlations \cite{Modi2012,Goold_2015,Landi2020a}. It is defined for a density matrix $\rho$, of $N$ parties, as
\begin{equation}
\label{eq:totalcorr}
\mathcal{T}(\rho) = \sum_{i=1}^{N}S(\rho_i)-S(\rho),
\end{equation}
where $\rho_i$ is the reduced density matrix of the i-th subsystem, and  $S(\rho)=-\mathrm{Tr} \left[ \rho\log{\rho} \right]$ is the von Neumann entropy of state $\rho$. The total correlation is thus positive by construction, and represents the total amount of information contained in the global state $\rho$, which is lost upon marginalization. It is worth pointing out that for bipartite systems, the total correlations equal the mutual information, defined as 
\begin{equation}
I_{\rho_{AB}}=S(\rho_A)+S(\rho_B)-S(\rho_{AB}),
\label{correlation}
\end{equation}
where $\rho_A=\mathrm{Tr_B}(\rho_{AB})$ and $\rho_B=\mathrm{Tr_A}(\rho_{AB})$.

For pure bipartite states, the total correlations are proportional to the entanglement of formation~\cite{Wootters1998,Horodecki2009}, defined as
\begin{equation}\label{eq:EoF}
E_{F}(\rho_{AB})=S(\rho_A)=-\mathrm{Tr}\left(\rho_A \log{\rho_A}\right),
\end{equation} 
and can be also understood as the mutual information between subsystems $A$ and $B$, given that $I(A:B)=S(\rho_A)+S(\rho_B)-S(\rho_{AB})=2 E_F(\rho_{AB})$ (since for pure states the von Neumann entropy is null). 

Notice that the global system-environment interaction can always be assumed to be unitary and thus cannot change the von Neumann entropy of the joint state, that is, $S(\rho^{\prime}_{SE})=S(\rho_{SE})$. Furthermore, since the system and environment are assumed to be initially uncorrelated, their initial mutual information $I_{\rho_{SE}}=S(\rho_{S})+S(\rho_{E})-S(\rho_{SE})$ is null and thus $S(\rho^{\prime}_{SE})=S(\rho_{SE})=S(\rho_{S})+S(\rho_{E})$. The mutual information $I_{\rho^{\prime}_{SE}}=S(\rho^{\prime}_S)+S(\rho^{\prime}_E)-S(\rho^{\prime}_{SE})$ at later times can then be rewritten as
\begin{equation}
I_{\rho^{\prime}_{SE}}=S(\rho^{\prime}_S)+S(\rho^{\prime}_E)-S(\rho_S)-S(\rho_E).
\end{equation}

Using the definition \eqref{eq:totalcorr} of total correlations we obtain
\begin{eqnarray}
\nonumber
I_{\rho^{\prime}_{SE}} & & =\sum_{i=1}^{N}\left\{S(\rho^{\prime}_{S_i})+S(\rho^{\prime}_{E_i})-S(\rho_{S_i})-S(\rho_{E_i})\right\}\\
& &-\mathcal{T}(\rho^{\prime}_S)+\mathcal{T}(\rho_S)+\mathcal{T}(\rho_E)-\mathcal{T}(\rho^{\prime}_E).
\end{eqnarray}
Note that the terms between brackets only involve local quantities and in particular define the local mutual information between subsystems and its respective environments:
\begin{eqnarray}
I^{\mathrm{local}}_{\rho^{\prime}_{SE}} & &=\sum_{i=1}^{N}I_{\rho^{\prime}_{S_i,E_i}}\\
& & =\sum_{i=1}^{N}\left\{S(\rho^{\prime}_{S_i})+S(\rho^{\prime}_{E_i})-S(\rho_{S_i})-S(\rho_{E_i})\right\}.
\nonumber
\end{eqnarray}
We then obtain that the variation $\Delta\mathcal{T}_S=\mathcal{T}(\rho^{\prime}_S)-\mathcal{T}(\rho_S)$  in the total correlation of the system can be generally decomposed as
\begin{equation}
\Delta \mathcal{T}_S=-I_{\rho^{\prime}_{SE}}+I^{\mathrm{local}}_{\rho^{\prime}_{SE}} -\Delta \mathcal{T}_E,
\end{equation}
where $\Delta \mathcal{T}_E=\mathcal{T}(\rho^{\prime}_E)-\mathcal{T}(\rho_E)$ is the corresponding variation of the total correlations in the environment. Interestingly, the equation above shows a rule for the flow of information between system and environment.

\begin{figure*} [!t]
\includegraphics[width=.98\textwidth]{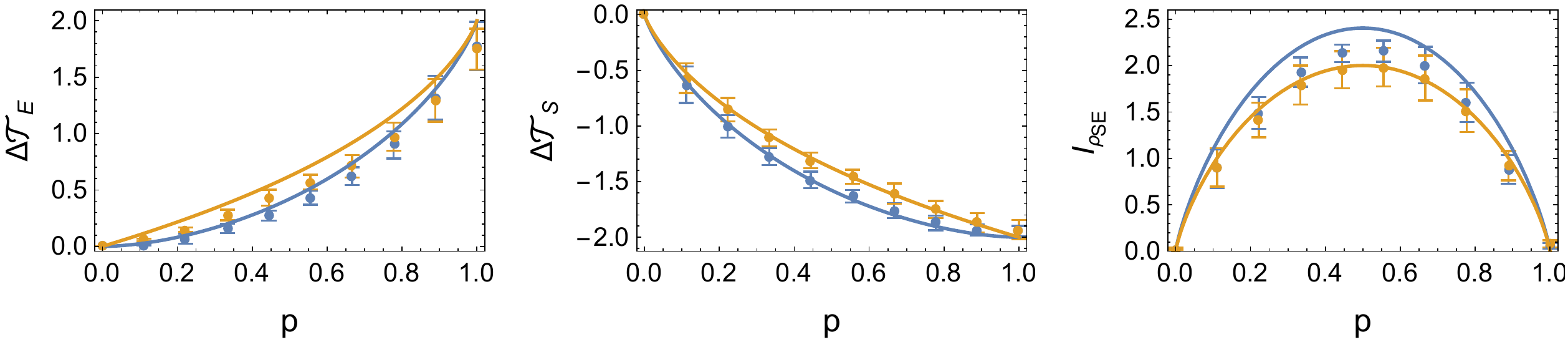}
\caption {\textbf{Dynamics of correlations for the amplitude damping dynamics.} Left:  increase in the total correlations of the environment. Center: decrease of total correlation of the system.
Right: mutual information built between system and environment. The system is initially in $\ket{01}+\ket{10}$ (yellow curve for theoretical result and yellow dots for experimental results) or $\ket{00}+\ket{11}$ (blue curve for theoretical result and blue dots for experimental results). All experimental results were obtained using the quantum computer ibmq\_manila from IBM Quantum Experience \cite{IBMQX, gadi2019}.}
\label{fig:plot1}
\end{figure*}

Naively, one might think that in order to minimize the loss of correlations within the system, it would be enough to simply minimize its correlations with the environment. The presence of the terms $-\Delta \mathcal{T}_E$ and $I^{\mathrm{local}}_{\rho^{\prime}_{SE}}$ shows, however, that the correlation dynamics is more intricate. As we will see below with specific examples, in a decoherent dynamics the total correlations of the system decrease with time, that is, $\Delta \mathcal{T}_S$ is negative. In turn, this will not only lead to correlations to be built between system and environment, be it the global $I_{\rho^{\prime}_{SE}}$ or the local $I^{\mathrm{local}}_{\rho^{\prime}_{SE}}$ mutual information, but also to a positive variation in total correlations between the different components of the environment.

\subsection{The effect of local unitaries}
The central question we pursue in this paper is to understand how local unitaries $U_i$, applied before the interaction with the baths, affects the correlations in the system. That is, we aim to compare the evolution of the correlations in the  state $\mathrm{Tr}_E\left(\rho^{\prime}_{SE}\right)= \Lambda(\rho(0))$ with those in 
$ \tilde{\rho}_S'= \Lambda((U_1\otimes \dots \otimes U_N)\rho(0)(U^{\dagger}_1\otimes \dots \otimes U^{\dagger}_N))$.

In general the unitaries optimizing the total correlations of the state over time will depend intrinsically on the initial state under consideration, and the type of noisy dynamics. In the particular case where the single qubit marginal states are the maximally mixed state, for instance, as it happens for bipartite maximally entangled states or the GHZ states, we notice that the individual states of the system remain unaltered by local unitaries. This implies, in particular, that the local correlations between system and environment $I^{\mathrm{local}}_{\rho^{\prime}_{SE}}$ are independent of the local unitaries. Hence, there will be a conservation law, given by
\begin{equation}
\Delta \mathcal{T}_S+I_{\rho^{\prime}_{SE}}+\Delta \mathcal{T}_E=\Delta \mathcal{T}_{\tilde{S}}+I_{\tilde{\rho}^{\prime}_{SE}}+\Delta \mathcal{T}_{\tilde{E}},
\end{equation}
where $\Delta \mathcal{T}_{\tilde{S}}=\mathcal{T}(\tilde{\rho}^{\prime}_S)-\mathcal{T}(\tilde{\rho}_S)$ and similarly for $\Delta \mathcal{T}_{\tilde{E}}$.

For states with maximally mixed marginals we thus see that the local correlations generated between system and environment play no role in the optimization provided by local unitaries.

\begin{figure*} [!t]
\includegraphics[width=.98\textwidth]{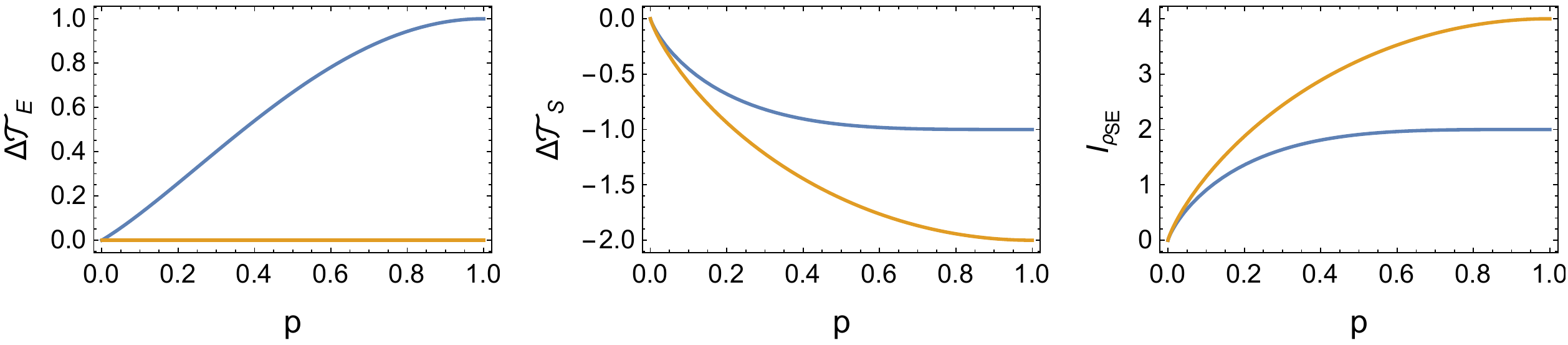}
\caption {\textbf{Dynamics of correlations for the dephasing dynamics.} Left: increase in the total correlations of the environment. Center: decrease of the total correlation of the system. Right:  mutual information built between system and environment. We consider that the system is initially in the quantum state $\ket{01}+\ket{10}$ (yellow curve) or in the state $\ket{0+}+\ket{1-}$ (blue curve).} \label{fig:plot2}
\end{figure*}

Figs. \ref{fig:plot1} and \ref{fig:plot2} consider specific examples showing the interplay and flow of these correlations during the system-environment evolution. In Fig. \ref{fig:plot1} the AD dynamics is analyzed by considering two different initial states: $\ket{01}+\ket{10}$ and $\ket{00}+\ket{11}$, both maximally entangled states related by simple local unitaries. As one could intuitively expect, the most robust state, that losing its total correlations slower than the others, is exactly that generating the fewer correlations with the environment. Curiously, however, is the fact that the more robust is the state the faster the total correlations built up in the environment. As shown in Fig. \ref{fig:plot2}, in which we consider the initial states  $\ket{01}+\ket{10}$ and $\ket{0+}+\ket{1-}$, a similar behaviour happens for the dephasing dynamics: the most robust state seems to be the one generating less correlations with the environment. In this case, however, we see that the initial state $\ket{0+}+\ket{1-}$ is such that the system loses its total correlations but no total correlations are transferred to the environment.

\section{Entanglement-dynamics in open quantum systems}
\label{sec:sec4}
Moving beyond the total correlations we also analyze the entanglement dynamics generate by the system and environment interaction. For pure bipartite states, the central entanglement quantifier is the entanglement of formation~\eqref{eq:EoF}.
For mixed states one should optimize over all possible decompositions of the quantum state $\rho_{AB}$. And, precisely for that aim, the concurrence  $\mathcal{C}$ has been introduced \cite{Wootters1998}. Not only it allows to compute the entanglement of formation for any bipartite qubit states, but has itself turned into an important entanglement monotone. It is defined for a pair of qubits as 
\begin{eqnarray}
\mathcal{C}[\rho] = \max\{0,\lambda_1 - \lambda_2 - \lambda_3 - \lambda_4\},
\label{concurrence}
\end{eqnarray}
in which $\{\lambda_1,\lambda_2,\lambda_3,\lambda_4\}$ are the singular values, in decreasing order, of the operator 
\begin{eqnarray}
\omega = \sqrt{\sqrt{\rho}(\sigma_y\otimes\sigma_y)\rho^*(\sigma_y\otimes\sigma_y)\sqrt{\rho}},
\end{eqnarray}
where $\sigma_y$ is a Pauli matrix, and $\rho^*$ ins the complex conjugate of $\rho$.

Important for our objectives, the concurrence can be expressed using a factorization law \cite{Konrad2008}. For instance, for a one-side channel it follows that 
\begin{equation}
\mathcal{C}\left[ \left( I\otimes \Lambda_2 \right) \rho \right] \leq \mathcal{C}\left[ \left(
I\otimes \Lambda_2 \right) \left\vert \Phi ^{+}\right\rangle \left\langle
\Phi ^{+}\right\vert \right] \mathcal{C}\left[ \rho \right] .
\label{facequal}
\end{equation}%
By the convexity of the concurrence, this can be turned into an inequality
for two-sided channels as
\begin{eqnarray}
& \mathcal{C}\left[ \left( \Lambda _{1}\otimes \Lambda _{2}\right) \rho \right]
\leq \label{facineq} \\ \nonumber
& \mathcal{C}\left[ \left( \Lambda _{1}\otimes I\right) \left\vert \Phi
^{+}\right\rangle \left\langle \Phi ^{+}\right\vert \right] \mathcal{C}\left[ \left(
I\otimes \Lambda _{2}\right) \left\vert \Phi ^{+}\right\rangle
\left\langle \Phi ^{+}\right\vert \right] \mathcal{C}\left[ \rho \right],
\end{eqnarray}%
with $\ket{\Phi ^{+}}=(1/\sqrt{2})(\ket{00}+\ket{11})$.

In turn, the negativity ($\mathcal{N}$) \cite{Vidal2002} is an easily computable measure of entanglement in any bipartition of multipartite states, being defined for a two qubit state as
\begin{eqnarray}
\mathcal{N}(\rho) = \frac{\norm{\rho^{T_A}}_1 - 1}{2},
\label{negativity}
\end{eqnarray}
$\rho^{T_A}$ standing for the partial transpose of $\rho$ in the bipartition under consideration. Importantly, the negativity upper bounds the singlet fraction \cite{Horodecki1999}, the maximal overlap of the state with a maximally entangled state $\ket{\psi}$, that is, $F(\rho)=\max \bra{\psi} \rho\ket{\psi}$, as $F(\rho)\leq \left[1+\mathcal{N}(\varrho)\right]/2$. In turn, $F$ is an operational entanglement measure, related to the teleportation fidelity \cite{Bennet1993, Verstraete2002} as $f=(2F+1)/3$.

\subsection{Optimal local encoding protecting the entanglement evolution}

A typical scenario we encounter in quantum information is given by a source which has to distribute entangled pairs among distant nodes. At the source one can generate high quality entangled pairs, that if assumed to be pure can generally be written as $(U_1 \otimes U_2) \ket{ \Psi _{\theta}}$ where $U_1$ and $U_2$ are local unitaries and
\begin{eqnarray}
\label{eq: Psi_theta}
\ket{ \Psi _{\theta}}= \cos \theta \ket{01} +\sin \theta \ket{10},
\end{eqnarray}
in which $\left\{\ket{0},\ket{1} \right\}$ is the computational basis representing the eigenstates of the Pauli matrix $\sigma_z$. Notice that local unitaries do not change the entanglement of the initial state $\ket{\Psi_{\theta}}$, that is, $E(\ket{\Psi_{\theta}})=E((U_1 \otimes U_2) \ket{ \Psi _{\theta}})$. And even though the local unitaries do not change the entanglement of the initial state, they might greatly affect the entanglement over the decoherent dynamics \cite{Hein2005,Ziman2006,cavalcanti2009open,Chaves2012}. Given that, it is natural to wonder what is the optimal local encoding (local unitaries) leading to the most robust entanglement over time.

In particular, given a certain noise channel, and an initial entangled state, one would like to find the most robust encoding (that is, the unitaries $U_1$ and $U_2$) preserving the most entanglement of the state through the noisy evolution.  Ideally, this enhanced robustness should hold to any entanglement measure $E$. Unfortunately this is not possible, since even in the bipartite case different entanglement measures do not generally agree on the ordering of states, that is, $E_1(\varrho_1) < E_1(\varrho_2)$ but possibly $E_2(\varrho_1) > E_2(\varrho_2)$. Because of that we will rely on the  particular measures of entanglement introduced above, such as the concurrence \cite{Wootters1998} and negativity \cite{Vidal2002}.

We will also be interested in quantifying the degree of entanglement between system and environment. The decoherence in the system of interest can always be understood via its unitary coupling to external degrees of freedom.  That is, assuming that initially both the system and environment are pure states, the whole system-environment undergoes a joint unitary transformation and it is maintained pure along all the evolution. In this case, to quantify the entanglement of formation $E_{SE}$ between system and environment we only need to compute the von Neumann entropy of the system, that is, $E_{SE}=S(\rho_{S})=-\rho_{S} \log{\rho_{S}}$.

To begin with, we consider the concurrence of the noisy states. Starting with the dephasing channel (D), it is straightforward to compute the concurrence if we set $U_1=U_2=I$, obtaining
\begin{equation}
\mathcal{C}\left( \left( \Lambda^D _{2}\otimes \Lambda_2^{D}\right) \left\vert
\Psi _{\theta}\right\rangle \left\langle \Psi _{\theta}\right\vert \right)  =
\left( 1-p_{1}\right) \left( 1-p_{2}\right) \sin \left( 2\theta \right),
\end{equation}
A direct comparison shows that the concurrence for state $\ket{ \Psi _{\theta} }$ undergoing local dephasing reaches the bound (\ref{facineq}). That is, $\ket{ \Psi _{\theta} }$ is the state with the most robust concurrence against local dephasing.

A similar analytical conclusion is obtained if we consider the Amplitude Damping channel (AD) or even for a concatenation of D and AD channels. We further note that since any local rotation $e^{i \phi Z}$ commutes with the channels D and AD, there are infinitely many local equivalent initial states with the same entanglement dynamics. Moreover, for the D channel, the state $\ket{ \Phi _{\theta}}=\cos \theta \ket{ 00} +\sin \theta \ket {11}$ also satisfies the bound (\ref{facineq}), and also provides the most robust concurrence.

Since we are interested in the relation between the robustness of state and the entanglement that it generates with the environment, more than knowing the most robust state, it is also useful to know how the  system-system entanglement evolution depends on the local unitaries. In this case the bound \eqref{facineq} is of no use and we must resort to the direct brute-force calculation of the concurrence, taking into account the local unitaries. Focusing on a a maximally entangled state, we can perform this calculation analytically. First, note that
\begin{eqnarray*}
\left\vert \Phi \right\rangle  &=&\left( U_{1}\otimes U_{2}\right)
\left\vert \Phi ^{+}\right\rangle =\left( I\otimes U\right) \left\vert \Phi
^{+}\right\rangle  \\
&=& \frac{1}{\sqrt{2}}  e^{i\phi_{1}Z_1+ i\phi_{2}Z_2 } \left( \left\vert 0\right\rangle \left\vert \Psi _{0}\right\rangle +\left\vert 1\right\rangle \left\vert \Psi _{1}\right\rangle \right),
\end{eqnarray*}
with%
\begin{eqnarray*}
\left\vert \psi^{\gamma} _{0}\right\rangle  &=&\cos \gamma \left\vert 0\right\rangle
+\sin \gamma \left\vert 1\right\rangle , \\
\bigskip \left\vert \psi^{\gamma} _{1}\right\rangle  &=&-\sin \gamma \left\vert
0\right\rangle +\cos \gamma \left\vert 1\right\rangle.
\end{eqnarray*}%
Since the local phases commute with the D and AD channels (or their concatenation), they cannot change the entanglement dynamics and the optimization can be restricted to states $\ket{\Phi_{\gamma}}=\left\vert 0\right\rangle \left\vert \psi^{\gamma} _{0}\right\rangle +\left\vert 1\right\rangle \left\vert \psi ^{\gamma}_{1}\right\rangle$. For the AD channel, the concurrence can be analytically computed and shown to be a monotonically decreasing function of $\gamma$, reaching its maximum at $\gamma=\pi/2$ and minimum at $\gamma=0$. That is, as argued previously, for the AD channel the maximally entangled state with most robust concurrence is $\ket{\Phi_{\pi/2}}=(1/\sqrt{2})(\ket{01}+\ket{10})$. In turn, the least robust is $\ket{\Phi_{0}}=(1/\sqrt{2})(\ket{00}+\ket{11})$.

Considering now the dephasing channel, the maximum is achieved for $\gamma= \left\{0,\pi/2 \right\}$ and the minimum for $\gamma=\pi/4$. As concluded previously, the state with the most robust concurrence against dephasing is $ \ket{\Phi_{\pi/2}}=(1/\sqrt{2})(\ket{01}+\ket{10})$. In turn, the least robust is $\ket{\Phi_{\pi/4}}=(1/\sqrt{2})(\ket{0+}+\ket{1-})$, corresponding to a two-qubit graph state \cite{Raussendorf2003} where $\ket{\pm}=(1/\sqrt{2})(\ket{0}\pm \ket{1})$.

In spite of having no factorization law~\eqref{facequal} for the negativity, we can verify by inspection what are the most and least robust maximally entangled states undergoing local noise. Considering the AD channel, the negativity is maximized for $\gamma=0$ and minimized for $\gamma=\pi /2$. Similarly, for the dephasing channel the negativity is maximized for $\gamma= \left\{0,\pi/2 \right\}$ and minimized for $\gamma=\pi/4$. That is, while for the D channel both concurrence and negativity agree on the ordering of states $\ket{\Phi_{\gamma}}$, for the AD channel the least robust maximally entangled state according to the negativity is the most robust according to the concurrence, and vice versa. 

\begin{figure} [!t]
\centering
\includegraphics[width=\linewidth]{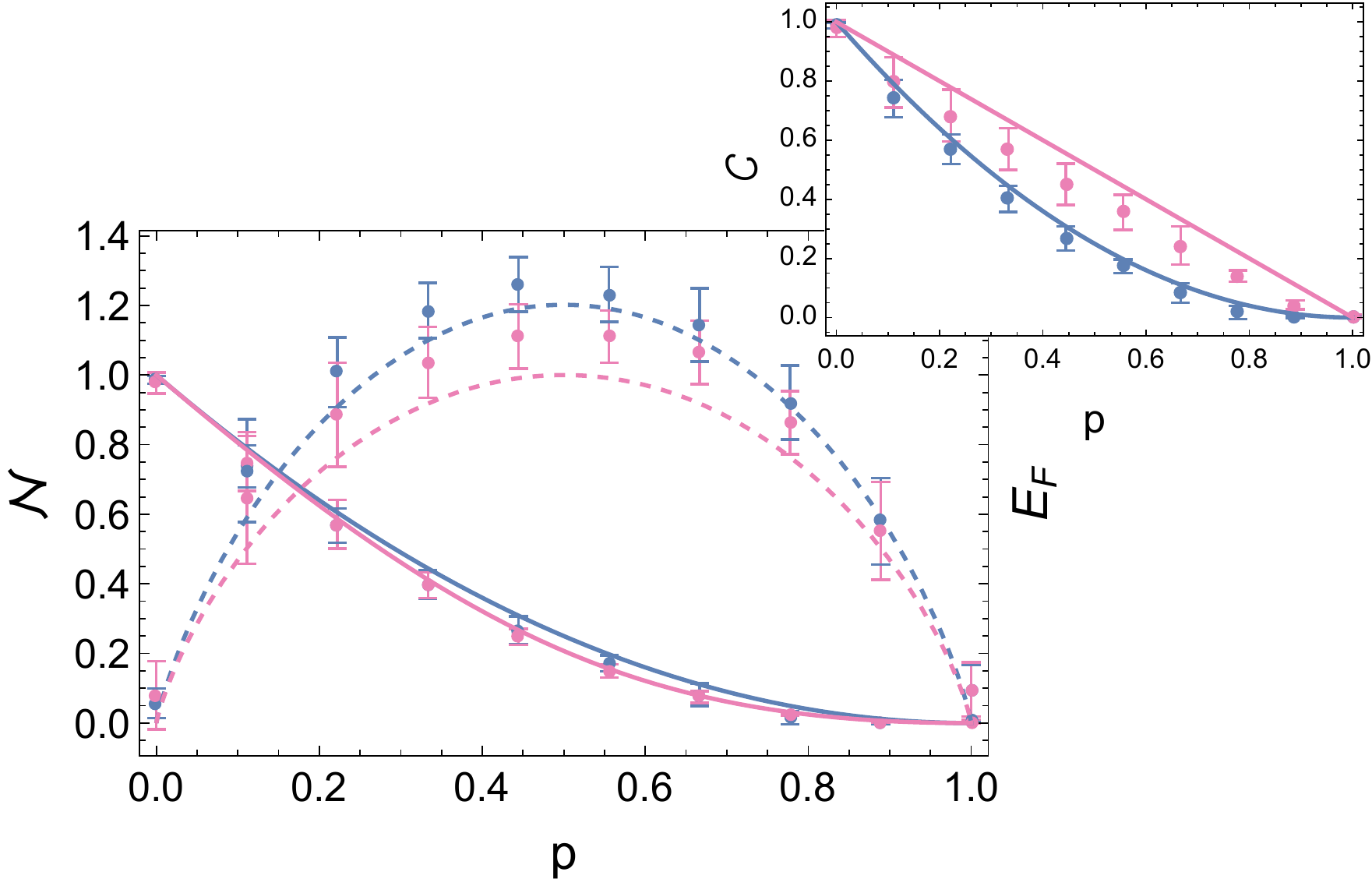}
\caption {The dynamics of $\mathcal{N}$, $\mathcal{C}$ (inset), $E_{E}(\varrho_{SE})$ for the maximally entangled state $\ket{\Phi_{\gamma=0}}$ (blue curves for theoretical results and blue dots for experimental results) and $\ket{\Phi_{\gamma=\pi/2}}$ (red curves for theoretical results and red dots for experimental results) undergoing the AD channel. Note that the ordering given by $\mathcal{N}$ and $\mathcal{C}$ is reversed. Surprisingly the most robust state according to the negativity is exactly the one generating more entanglement with the environment. All experimental results were obtained using the quantum computer ibmq\_manila from IBM Quantum Experience \cite{gadi2019}}
\label{fig:neg_ad_entropy}
\end{figure}

We turn our attention to the entanglement generated between the system and its environment along the noisy evolution. Once more, we restrict to the maximally entangled states. It is not difficult to compute the eigenvalues of the general state $\ket{\Phi_{\gamma}}$ undergoing AD or D channels. It can be seen that for the D evolution, for both the concurrence and the negativity, the more robust is the state, the less entanglement is created with the environment. However, for the AD channel the situation changes. As shown in  Fig. \ref{fig:neg_ad_entropy}, while for the concurrence the expected intuition prevails, for the negativity, the most robust state is exactly the one maximizing the entanglement with the environment. Remarkably, for $p_1=p_2=p$, its negativity reads,
\begin{eqnarray}
\mathcal{N} = 1-2p+p^2
\end{eqnarray}
which equals to $2F(\rho_{\gamma =0}) - 1$, $F(\cdot)$ being the singlet fraction and $\rho_{\gamma =0}$ the state of the system after the AD channel. Hence, for those states the singlet fraction saturates the upper bound imposed by the negativity, which together with the fact that $\mathcal{N}(\rho_{\gamma=0})\geq\mathcal{N}(\rho_{\gamma=\pi/2})$, implies that even though $\ket{\Phi_{\gamma=0}}$ generates more entangled with the environment, nonetheless is also the most useful state in a teleportation protocol.

\subsection{The maximally entangled state is not the optimal resource in the noisy scenario.}
In an ideal noiseless scenario the maximally entangled state is the paradigmatic state, allowing the optimal execution of information tasks. For one side channels it is known that the maximally entangled states are still the most robust states, maximizing the entanglement over the whole dynamics \cite{Ziman2006}. Formally, for any state
$\rho $,
\begin{equation}
E\left( \left( I\otimes \varepsilon \right) \rho \right) \leq E\left( \left(
I\otimes \varepsilon \right) \left\vert \Psi \right\rangle \left\langle \Psi
\right\vert \right) ,
\label{oneside}
\end{equation}%
where $E$ is any entanglement monotone and $\left\vert \Psi \right\rangle $ represents some maximally entangled state. As we show next, the maximally entangled state is not necessarily the most robust against noise for the more realistic scenario of two-side channels. Moreover, considering a teleportation protocol, this leads to an unexpected situation, where in the noisy scenario it is advantageous to start out with a less entangled state than a maximally entangled one.

\begin{figure} [!t]
\centering
\includegraphics[width=\linewidth]{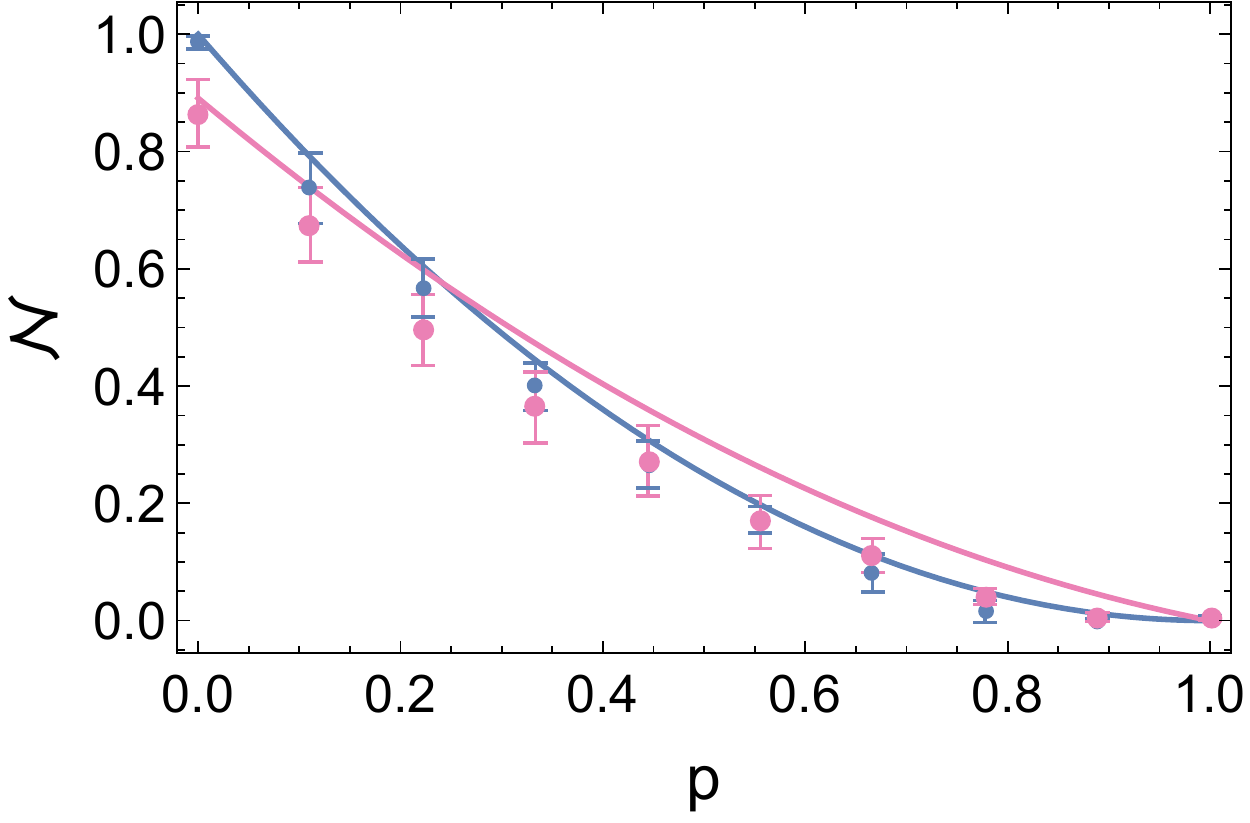}
\caption {Negativity dynamics for the AD channel. The blue curve and dots represents the maximally entangled state while the red curve and dots the non-maximally entangled state.(with $\theta=0.7\frac{\pi}{4}$ for example), curve for theorical results and dots for experimental results} \label{fig:crossing}
\end{figure}

To prove that (\ref{oneside}) is not valid in general we will focus on the negativity \cite{Vidal2002} and the AD channel. As shown before, the most robust maximally entangled state according to the negativity is $ \ket{\Phi_{\gamma=0}}$. Plotting the negativity for $ \ket{\Phi_{\gamma=0}}$ and the negativity for a non-maximally entangled state $ \ket{\Psi_{\theta}}$, following definition \eqref{eq: Psi_theta}, for a certain $\theta \neq \pi/4$, one can see in Fig. \ref{fig:crossing} that the curves cross each other. That is, the initially more entangled state is not the most robust against noise along all the dynamics. Moreover, for the state $\ket{\Psi_{\theta}}$ undergoing local AD channels, it is possible to show that, for the resulting noisy state $\varrho_{\theta}$, $F(\varrho_{\theta})=\bra{\psi_{\theta}}\varrho_{\theta}\ket{\psi_{\theta}}=(1+\mathcal{N}(\varrho_{\theta}))/2$. That is, in this particular case the negativity gains an operational meaning in terms of the teleportation fidelity. Thus showing that a less entangled state may be a better resource in the realistic implementation of some quantum information tasks.

\section{Discussion}
\label{sec:sec5}
To counteract the unavoidable effects of decoherence is a step of primal importance for the establishment of quantum technologies. Error correction provides a general framework for that, however, due to the large overhead in extra qubits, it remains as a impractical solution to near term quantum devices. It is thus important to explore alternative and experimental less demanding ways to preserve the quantum properties of state and consequently improve its use as resource.

Previously, considering particular classes of multipartite states, it has been noticed that simple local unitaries can greatly improve the robustness of the entanglement dynamics \cite{Chaves2012}. However, even for bipartite systems, it was not known what are the unitaries leading to an optimal dynamics.

Here we solve this question by analyzing two paradigmatic and relevant noise channels, the amplitude damping and dephasing and combinations of both. For pure initial qubit states we analytically find what are the unitaries leading to best and worst concurrence. As one could expect, the more the system is entangled with the environment, the lower is the concurrence between the qubits of the system. As a proof-of-principle, we also show the effect of these unitaries in practice, by implementing the corresponding quantum circuit simulating the the system-environment evolution in the IBM Quantum experience.

By considering the negativity instead, a surprisingly effect occurs: the optimal unitary is exactly that generating the most entanglement with environment. Furthermore, we show that the negativity of a initially less entangled state can be more robust against decoherence than a maximally entangled one. As a consequence, there is a range of noise parameters for which the best teleportation protocol is achieved with such an initially less entangled state. The fact that to start out with a less entangled state provides a larger probability of success is a a result that resembles the seminal result of Eberhard \cite{Eberhard1993}, in the achievement of a loophole free Bell test.

Finally, we go beyond the understand of entanglement dynamics and also derive a general law relating the mutual information between system and environment with the changes in the total correlation. Even thought we cannot reach general conclusions, for the specific cases we have analyzed the unitary that best preserves the total correlations of the system is exactly that generating less mutual information/entanglement between system and environment. At the same time, however, this optimal unitary is the one that generates the most correlations between the different parts of the subsystem. 

We believe our results may help to enhance the robustness of quantum information protocols of practical relevance and to better understand the dynamics of quantum correlations and their use as a potential resource. In particular, it would be interesting to understand whether local unitary encodings can also lead to improvements in quantum computation algorithms. We hope our results might trigger further developments in this area.

\begin{acknowledgments}
We acknowledge the use of IBM Quantum services for this work. The views expressed are those of the authors, and do not reflect the official policy or position of IBM or the IBM Quantum team. This work was supported by the John Templeton Foundation via the grant Q-CAUSAL No 61084 (the opinions expressed in this publication are those of the author(s) and do not necessarily reflect the views of the John Templeton Foundation) Grant Agreement No. 61466,  by the Serrapilheira Institute (grant number Serra – 1708-15763), by the Simons Foundation (Grant Number 884966, AF), the Brazilian National Council for Scientific and Technological Development (CNPq) via the National Institute for Science and Technology on Quantum Information (INCT-IQ) and Grants No. 406574/2018-9 and 307295/2020-6, the Brazilian agencies MCTIC and MEC.
\end{acknowledgments}

\end{document}